\documentclass{emulateapj}
\usepackage{apjfonts}
\usepackage{epsfig}

\def\alphasn{\alpha_{_{\rm SN}}}

\def\calg{{\cal G}}
\def\calr{\dot{{\cal R}}_*}
\def\civ{C~{\sc iv}}
\def\dotm{\dot{M}}
\def\dotmbh{\dot{M}_{\bullet}}

\def\dotcalms{\dot{\cal M}_*}
\def\dotmout{\dot{M}_{\rm out}}
\def\dotrhos{\dot{\rho}_*}

\def\mbh{M_{\bullet}}
\def\mz{m_{_{Z}}}
\def\nusn{\nu_{_{\rm SN}}}
\def\nv{N~{\sc v}}
\def\rhog{\rho_{\rm gas}}
\def\siggas{\Sigma_{\rm gas}}
\def\sigsfr{\dot{\Sigma}_*}
\def\sunm{M_{\odot}}
\def\sunmyr{M_\odot{\rm yr^{-1}}}

\journalinfo{The Astrophysical Journal Letters, 719: (in press)}
\slugcomment{Received 2010 April 14; accepted 2010 July 14}
\shorttitle{Accretion disk in AGNs is driven by star formation}

\begin{document}

\title{Accretion Disks in Active Galactic Nuclei: Gas Supply Driven by Star Formation}

\author{Jian-Min Wang, Chang-Shuo Yan, Han-Qin Gao, Chen Hu, Yan-Rong Li and Shu Zhang}

\affil{\small
Key Laboratory for Particle Astrophysics, Institute of High Energy Physics,
Chinese Academy of Sciences, \\
19B Yuquan Road, Beijing 100049, China
}

\begin{abstract}
Self-gravitating accretion disks collapse to star-forming(SF) regions extending to the
inner edge of the dusty torus in active galactic nuclei (AGNs). A full set of equations
including feedback of star formation is given to describe the dynamics of the regions.
We explore the role of supernovae
explosion (SNexp), acting to excite turbulent viscosity, in the transportation of angular
momentum in the regions within 1pc scale. We find that
accretion disks with typical rates in AGNs can be driven by SNexp in the regions and
metals are produced spontaneously. The present model predicts a metallicity--luminosity
relationship consistent with that observed in AGNs. As relics of SF regions,
a ring (or belt) consisting of old stars remains for every episode of supermassive black
hole activity. We suggest that multiple stellar rings with random directions interact and form
 a nuclear star cluster after episodes driven by star formation.
\end{abstract}
\keywords{accretion disks - quasars: general - stars: formation}

\section{Introduction}
Accretion onto supermassive black holes (SMBHs) is the energy house of active galactic nuclei 
(AGNs). Applying
models of accretion disks to fit the observed big blue bumps in the continuum of AGNs and
quasars, one finds that SMBHs accrete typically with sub-Eddington rates (Laor
\& Netzer 1989; Sun \& Malkan 1989; Brunner et al. 1997; L\"u 2008). Observational
examination of the Soltan argument for the cosmological growth of SMBHs
prefers that they grow episodically through baryon accretion (Yu \&
Tremaine 2002; Wang et al. 2009a), in agreement with the paradigm of coevolution
with their host galaxies. However, what is governing accretion rates of active SMBHs?

We have several relevant clues to understanding the formation of the accretion flows
approaching SMBHs at $\sim 1$pc scale. First, the well-known metallicity and luminosity
relationship in AGNs and quasars (see a review of Hamann \& Ferland 1999) strongly
implies that the flows may undergo {\em fast} metal productions through stellar evolution
in active episodes in light of the absence of cosmic evolution of metallicity (Dietrich
et al. 2003; Shemmer et al. 2004). This gets supports from evidence
that there is intensive star formation with the  top-heavy initial mass function
(IMF) within $\sim 0.1$pc in the Galactic center (Paumard et al. 2006;
Nayakshin \& Sunyaev 2005) confirmed by numerical simulations (Nayakshin et al. 2007;
Bonnell \& Rice 2008; Hobbs \& Nayakshin 2009). We may naturally postulate that accretion
onto SMBHs is the consequence of metal production.
Second, it has long been known that accretion disks in quasars are massive and
self-gravitating in outer regions (Paczynski 1978; Kolykhalov \& Sunyaev 1980;
Shlosman \& Begelman 1987; Collin \& Zahn 1999). Obviously, the self-gravity is driving
intensive star formation ongoing there (Goodman 2003), and a star-forming (SF) disk (or 
regions) appears.
It seems that there is no way to avoid the production of metals and trigger accretion
flows in the SF disk, yielding the metallicity--luminosity relationship. We argue
that SNexp  plays a leading role in the establishment of the relationship in light
of SNexp-excited turbulent viscosity as shown in the $\sim 1$kpc regions by numerical
simulations (Wada \& Norman 2002; Hobbs et al. 2010) or physical motivation (Wang et
al. 2009b; see also Kumar \& Johnson 2010)
and $\sim 10^4$ Sloan type 2 AGNs (Chen et al. 2009). Furthermore, we are inspired by the
recent evidence for that the "visible" inflows displayed by the H$\beta$ line are likely
developed from the inner edge of the dusty torus in $\sim$ 5000 Sloan quasars (Hu et al.
2008a,2008b). All clues merge into a question: are accretion inflows driven by star formation
in the massive self-gravitating disk?

In this Letter, we show the feasibility that an SF disk is able to drive a Shakura--Sunyaev
disk (hereafter SS disk) in AGNs through SNexp-excited turbulent viscosity. We
find that such a scenario agrees with observations of metallicity and results in the formation
of a nuclear star cluster potentially.

{\centering
\figurenum{1}
\includegraphics[width=1.8in,angle=-90]{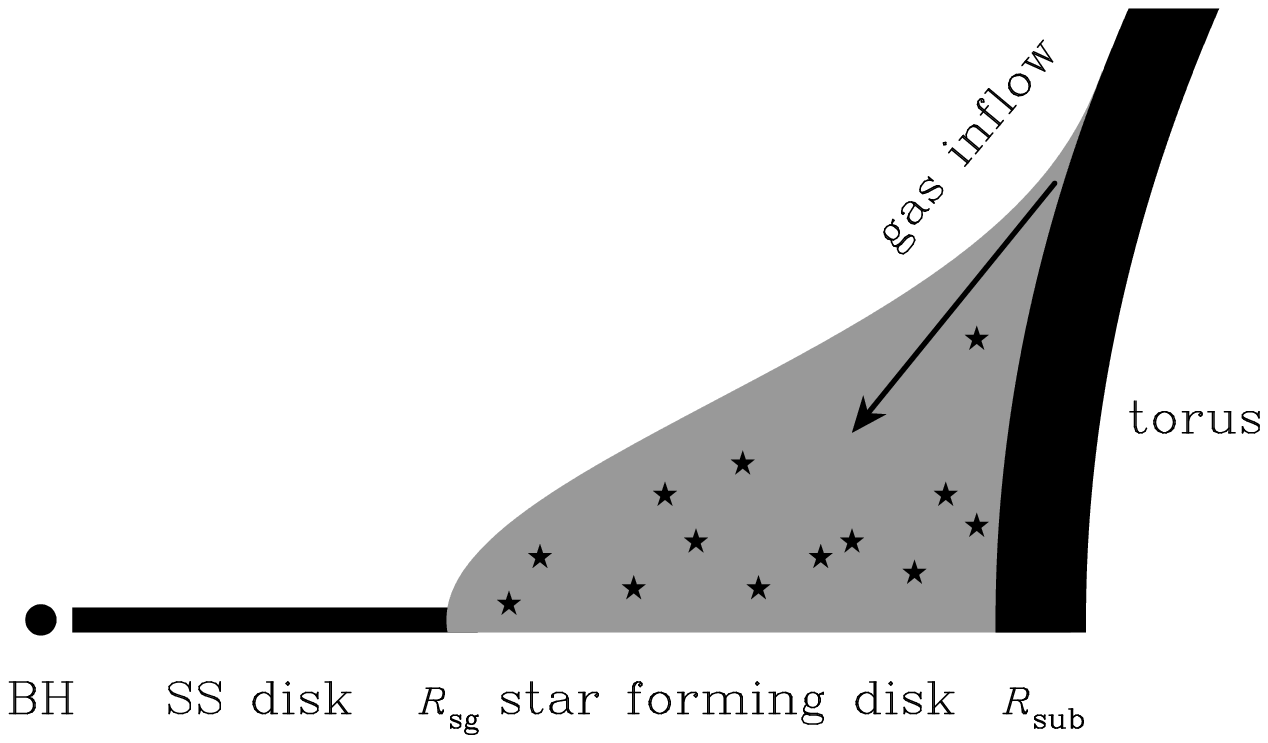}
\figcaption{Schematic illustration of the present model. The SF disk starts from the
inner edge of the dusty torus ($R_{\rm sub}$) and switches on the SS disk
at the self-gravity radius ($R_{\rm sg}$).}
}
\vglue 0.2cm

\section{Basic equations}
Figure 1 shows an illustration of the present model. The outer boundary of the SF disk
($R_{\rm out}$) is chosen at the inner edge of the torus as suggested by Hu et al. (2008a,2008b),
where dust particles are sublimated at $R_{\rm sub}=1.3L_{\rm UV,46}^{1/2}$ pc, where
$L_{\rm UV,46}=L_{\rm UV}/10^{46}$ erg~s$^{-1}$ (Barvainis 1987). The option of
$R_{\rm out}=R_{\rm sub}$ is motivated by a fact that emission from $R\ge R_{\rm sub}$
is invisible in optics. Collisions of molecular clouds transfer angular momentum
outward and supply gas to the SF disk with a rate of $\dotmout=1\sim 20\sunmyr$
at $R_{\rm sub}$ from the torus (Krolik \& Begelman 1988) with a typical mass range of
$10^5\sim 10^7\sunm$ (Mor et al. 2009). This typical mass is just
at the level required by a single episode of SMBH activity.

\subsection{SNexp-driven gaseous disk}
With the inclusion of the mass dropout from star formation and the injection of gas recycled
from SNexp, the mass conservation equation for the SF disk reads\footnote{Equation (1) is
the modified version from Lin \& Pringle (1987) and Wang et al. (2009b).
We point out that $\dot{\Sigma}_{\rm SN}$
is at $t-\tau_*$, where $\tau_*$ is the hydrogen main-sequence lifetime of stars
depending on the stellar mass. Since stars form in the same gaseous disk, the stellar disk always
follows the latter in the current model. Unless the stellar disk was formed in the previous episode,
it could be larger than
the gaseous disk, leading to reduce the feedback efficiency significantly (Nayakshin et al. 2007).}
\begin{equation}
\frac{\partial{\siggas}}{\partial{t}}=\frac{1}{2\pi R}\frac{\partial{}}{\partial{R}}
     \left\{\left[\frac{d \left(R^2\Omega\right)}{dR}\right]^{-1}
     \frac{\partial \calg}{\partial R}\right\}-\sigsfr+\dot{\Sigma}_{\rm SN},
\end{equation}

\noindent
where $\siggas$ is the surface density, $\calg=-2\pi R^3\nu\siggas\left(d\Omega/dR\right)$
is the viscosity torque,
$\nu$ is the kinematic viscosity, $\Omega$ is the angular velocity,
$\sigsfr$ is the surface density of star formation rates, and $\dot{\Sigma}_{\rm SN}$ is
the injection rate of gas recycled from SNexp. We use the relation $\dot{\Sigma}_{\rm SN}=f_c\sigsfr$, 
where $f_c$ is the fraction of the SNexp-ejected mass
to the total formed stars. For an episodic activity of AGN with $\Delta t_{\rm G}$ (e.g.,
Wang et al. 2006), the minimum mass of the stars producing an SNexp
should be $M_{\rm c}/\sunm\ge 7.0~\Delta t_{0.1}^{-0.4}$, where the lifetime of hydrogen
main-sequence stars is $t_{\rm MS}=13\left(M_*/\sunm\right)^{-2.5}$Gyr and
$\Delta t_{0.1}=\Delta t_{\rm G}/0.1{\rm Gyr}$. Assuming an IMF as
$N(M_*)dM_*=N_0 M_*^{-\beta}dM_*$, we have
$f_c=\int_{M_c}^{100\sunm} M_* N(M_*)dM_*/M_{\rm tot}$, where
$M_{\rm tot}=\int_{M_{\rm min}}^{100\sunm}M_*N(M_*)dM_*$ is the total mass of the formed
stars. It follows $f_c\sim 0.38$ for the Salpeter IMF $\beta=2.35$, $M_{\rm min}=1.0\sunm$,
and $M_c=7\sunm$, meaning that a significant mass fraction of the formed stars
will be injected into the SF disk from SNexp (here we neglect the remnant compact stars
after SNexp). The scale influenced by an SNexp is of the order of
$D_{\rm SN}\sim \left(E_{\rm SN}/m_pnc_s^2\right)^{1/3}
\sim 0.1~E_{51}^{1/3}\left(n_{11}T_3\right)^{-1/3}$ pc, where $E_{51}=E_{\rm SN}/10^{51}{\rm erg}$
is the kinetic energy of SNexp, $n_{11}=n_{\rm gas}/10^{11}{\rm cm^{-3}}$ is the number density
of the medium in the SF disk, $c_s\approx 3.0\times 10^5T_3^{1/2}$cm~s$^{-1}$ is the isothermal
sound speed, and $T_3=T/10^3$K is the temperature. We find from this simple estimation that the
length of the SNexp influence is comparable with the height of the SF disk ($\sim 0.1$ pc) and
the SNexp-excited turbulence velocity $V_{\rm tur}\gg c_s$, and hence SNexp plays a key role
in the transportation of angular momentum.

In this Letter, we focus on the stationary case, namely,
$\partial\siggas/\partial t=0$. Introducing the parameter $X=R^2\Omega$, Equation (1) is
rewritten as
\begin{equation}
\frac{1}{2\pi R}\frac{dX}{dR} \frac{d^2 \calg}{d X^2}-A(1-f_c)\siggas^{\gamma}=0,
\end{equation}
where the Kennicutt--Schmidt (KS) law $\sigsfr=A\siggas^{\gamma}$ is used for simplicity
in the entire SF disk, $A$ is a constant, and $\gamma$ is the index
(Kennicutt 1998)\footnote{We note that star formation rates may deviate from the KS law
along the radii in the SF disk. It is not sufficiently understood (Krumholz et al. 2009),
especially for so dense disks.}. Lynden-Bell \& Pringle (1974) show that
$d\calg/dX=\dot{M}$ at any radius, where $\dot{M}$ is the mass rate of the inflows,
provides the inner and outer boundary conditions.
We employ the $\alpha$-prescription viscosity
as $\nusn=\alphasn V_{\rm tur}H$, where $\alphasn$ is a constant, $V_{\rm tur}$ is the
turbulence velocity driven by SNexp, and $H$ is the thickness of the disk (Wada \& Norman
2002; Wang et al. 2009b). The energy equation of the SNexp excited turbulence is given by
\begin{equation}
\frac{\rhog V_{\rm tur}^2}{t_{\rm dis}}=\frac{\rhog V_{\rm tur}^3}{H}
            =\epsilon f_*\dotrhos E_{\rm SN},
\end{equation}
where $t_{\rm dis}=H/V_{\rm tur}$ is the timescale of the turbulence, $\epsilon$ is the
efficiency of converting kinetic energy of SNexp into turbulence, and $f_*\dotrhos$ is
the density of SNexp rates. Here $f_*$ is a number fraction of stars to the total mass
of the formed stars, which are able to produce SNexp during one AGN episode. For a
given IMF, we have $f_*=\int_{M_c}^{100\sunm}N(M_*)dM_*/M_{\rm tot}$ and
$f_*\sim 2.3\times 10^{-2}\sunm^{-1}$ for $\beta=2.35$. Considering $\sigsfr=\dotrhos H$
and $\siggas=\rhog H$, we have
\begin{equation}
\frac{V_{\rm tur}^3}{H}=\epsilon f_*AE_{\rm SN}\siggas^{\gamma-1}.
\end{equation}
Following Paczynski (1978), we assume that gas rotates with a Keplerian velocity
$\Omega_{\rm K}=\left(G\mbh/R^3\right)^{1/2}$, where $G$ is the gravity constant and
$\mbh$ is the SMBH mass. A vertical
equilibrium holds among the disk self-gravity, SMBH gravity, and turbulent pressure.
We usually have $dP_{\rm tur}/dz=\left(\Omega^2z+2\pi G\siggas\right)\rhog$, where
$P_{\rm tur}=\rhog V_{\rm tur}^2$ is the turbulence pressure and the second term is
the self-gravity. Since the star formation supports the thickness
of the SF disk, the self-gravity can be neglected (Thompson et al. 2005), allowing us
to have an approximation form of vertical-averaged structure,
\begin{equation}
\frac{P_{\rm tur}}{H}=\left(\frac{G\mbh}{R^3}H+2\pi G \siggas\right)\rhog
                     \approx \Omega_{\rm K}^2H\rhog.
\end{equation}
We point out that the above equations are actually averaged in the vertical direction,
and the parameters obtained in these equations are the values at the mid-plane of
the SF disk.

The $\calg-\siggas$ relation is specified for further simplification of Equation (2)
through the viscosity $\nu$. With the help of Equations (3)-(5), we have
\begin{equation}
\frac{d^2\calg}{dX^2}-\left(\frac{B}{X^2}\right)^2\calg=0,
\end{equation}
yielding an analytical solution in the form of
\begin{equation}
\calg(X)=c_1X\exp\left(\frac{B}{X}\right)+c_2X\exp\left(-\frac{B}{X}\right),
\end{equation}
where $B=2G\mbh\left[(1-f_c)/3\alpha_{\rm SN}\epsilon f_* E_{\rm SN}\right]^{1/2}$,
$X=\left(G\mbh R\right)^{1/2}$ for the Keplerian rotation, and $c_1$ and $c_2$ are two
constants determined by the boundary conditions. At the outer radius of the SF
disk, only the mass injection rate is known and given by the mechanism of molecular cloud
collisions (Krolik \& Begelman 1988). We have
$\left.{d\calg}/{dX}\right|_{X_{\rm out}}=\dot{M}_{\rm out}$, namely,
\begin{equation}
c_1\left(1-B X_{\rm out}^{-1}\right)e^{\frac{B}{X_{\rm out}}}+
c_2\left(1+B X_{\rm out}^{-1}\right)e^{-\frac{B}{X_{\rm out}}}=\dot{M}_{\rm out},
\end{equation}
where $X_{\rm out}=\left(G\mbh R_{\rm out}\right)^{1/2}$. We assume that the inner
radius of the SF disk is set to be the self-gravity radius ($R_{\rm sg}$) of the SS
disk, which depends on the SMBH accretion rates ($\dotmbh$). This yields
$\left.{d\calg}/{dX}\right|_{X_{\rm in}}=\dot{M}_{\bullet}$, and subsequently
\begin{equation}
c_1\left(1-B X_{\rm sg}^{-1}\right)e^{\frac{B}{X_{\rm sg}}}+
c_2\left(1+B X_{\rm sg}^{-1}\right)e^{-\frac{B}{X_{\rm sg}}}=\dot{M}_{\bullet},
\end{equation}
where
$X_{\rm in}=X_{\rm sg}=\left(G\mbh R_{\rm sg}\right)^{1/2}$.

{\centering
\figurenum{2}
\includegraphics[width=3.0in,angle=-90]{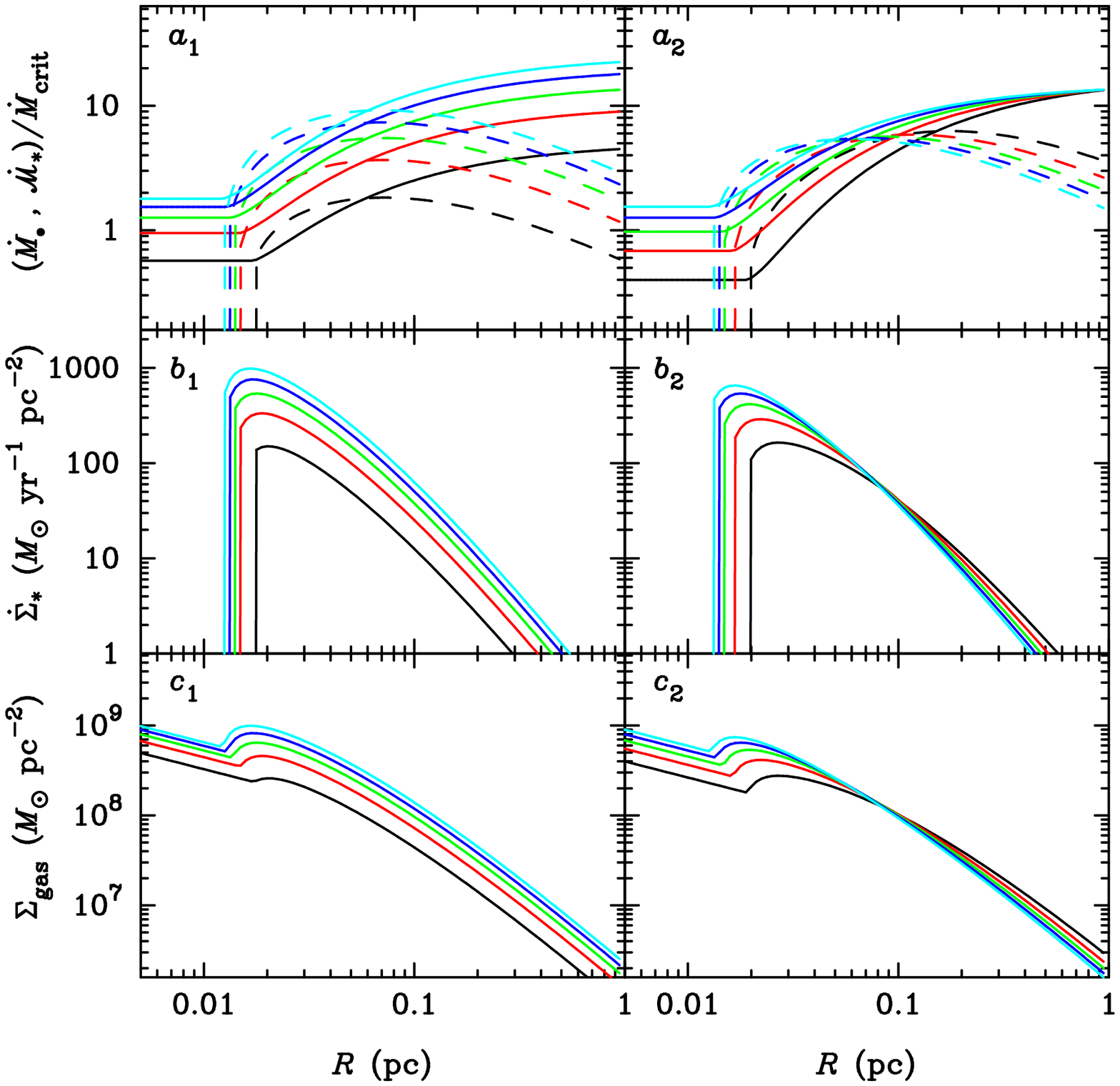}
\figcaption{Structure of SF disks, which drive SS
disks with an accretion rate of $0.5-2L_{\rm Edd}/c^2$, as shown by the solid
lines in $a_1$. The panels on the left column are for $\epsilon=0.5$, but $\dotmout$
takes 1.0, 2.0, 3.0, 4.0, 5.0$\sunmyr$ represented by the black, red, green, blue, and cyan
lines, respectively. The dashed lines are the local star formation rates defined by
$\dot{\cal M}_*=\pi R^2\sigsfr$, which are of $1 - 10\sunmyr$.
The panels on the right column are the structures for fixed
$\dotmout=3.0\sunmyr$, but $\epsilon=$ 0.2, 0.3, 0.4, 0.5, 0.6 by black, red, green,
blue, and cyan lines, respectively.}
}

\subsection{Switching on the Shakura--Sunyaev disk}
The self-gravity instability develops when the Toomre's parameter $Q\le 1$. For the SS
disk, the instability happens in the middle region of the SS disk at the radius
\begin{equation}
R_{\rm sg}=1068~\alpha_{0.1}^{14/27}M_8^{-26/27}\dot{m}^{-8/27}~R_{\rm Sch},
\end{equation}
where $R_{\rm Sch}=2G\mbh/c^2$ is the Schwartzchild radius, $c$ is the light speed,
$M_8=\mbh/10^8\sunm$,
$\dot{m}=\dotmbh c^2/L_{\rm Edd}=\dotmbh/0.2M_8\sunmyr$, $L_{\rm Edd}$ is the Eddington
luminosity, and
$\alpha_{0.1}=\alpha_{\rm SS}/0.1$ is the viscosity parameter in the SS disk (see Kato
et al. 1998). We assume that the SF process quenches within $R_{\rm sg}$, and the SF disks
smoothly switch on the SS disks at $R_{\rm sg}$ in term of surface density.
The surface density of the SS disks is
\begin{equation}
\Sigma_{\rm SS}=8.2\times 10^8~\alpha^{-4/5}M_8^{1/5}\dot{m}^{3/5}r_3^{-3/5}~{\sunm~\rm pc^{-2}},
\end{equation}
where $r_3=R/10^3R_{\rm Sch}$, giving rise to the expression of the condition
($\Sigma_{\rm gas}=\Sigma_{\rm SS}$) of switching from the SF regions to the SS disk
\begin{equation}
c_1e^{\frac{B}{X_{\rm sg}}}+c_2e^{-\frac{B}{X_{\rm sg}}}
=3\pi\alpha \epsilon f_* AE_{\rm SN}\Sigma_{\rm SS}^{\gamma}\Omega^{-2}.
\end{equation}
It is trivial to get the prolix expression of $c_1$ and $c_2$ from Equations (9) and
(12), but we omit them here. We insert $c_1$ and $c_2$ into Equation (8) for the
structure of the SF disk. Given $R_{\rm out}$ and $\dot{M}_{\rm out}$, we can justify
if the SF disk is able to drive the SS disk through solving equations (8), (9), (12).
The accretion rates of the SS disk as well as the structure
of the SF disk are obtained self-consistently.

{\centering
\figurenum{3}
\includegraphics[width=2.3in,angle=-90]{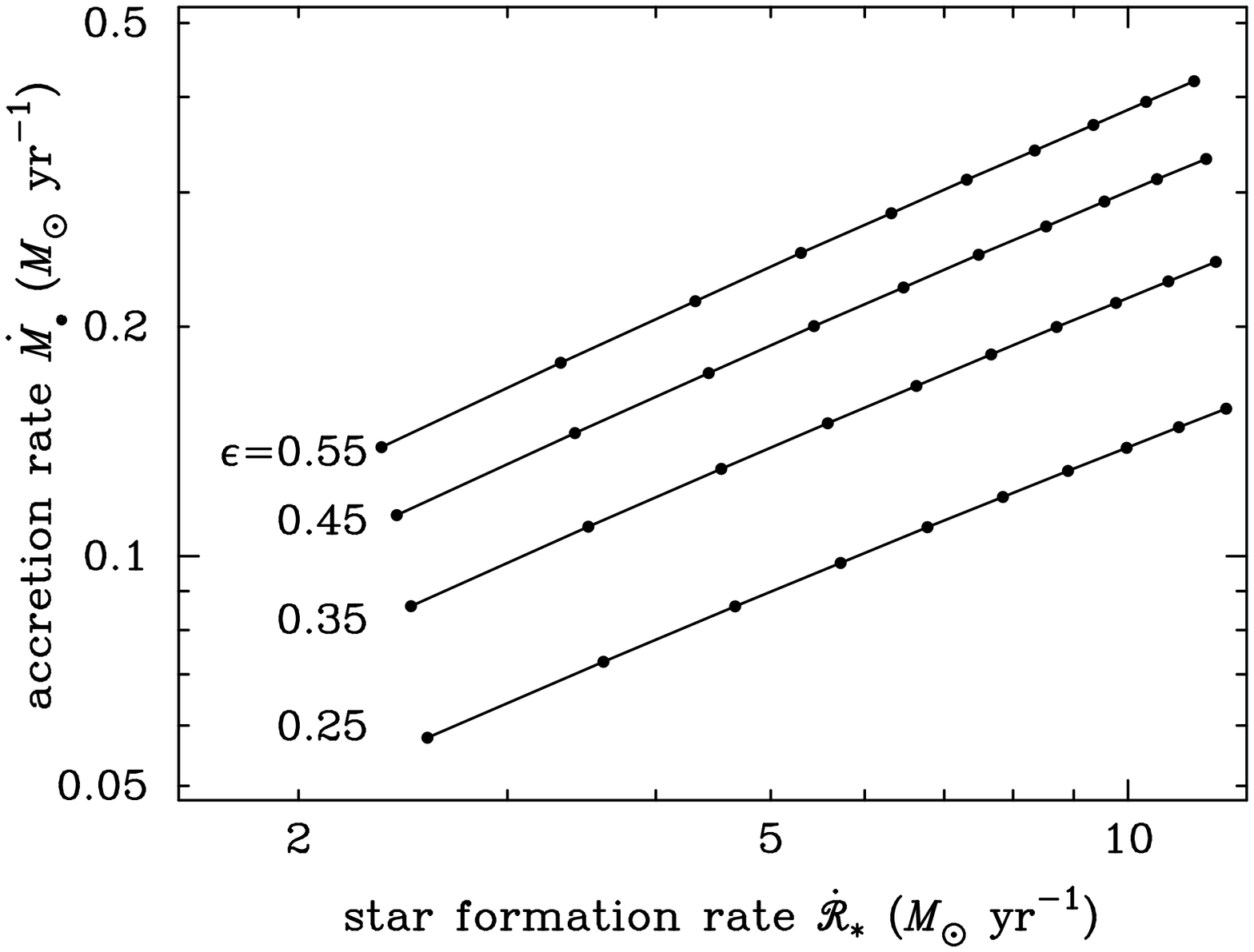}
\figcaption{Relation between black hole accretion rates and the star
formation rates in the SF  disks. Each line represents a relation for the labeled
$\epsilon$. We find that a relation of $\calr\propto \dotmbh^{1.4}$ and $\epsilon$ determines
intercepts of the relation.}
}

\section{Accretion flows approaching to SMBHs from SF disks}
The empirical KS law $\sigsfr=A\siggas^{\gamma}$ is used, where $\gamma=1.4$,
$A=2.5 \times 10^{-4}$, $\sigsfr$ and $\siggas$ are in units of
$\sunm \rm{yr}^{-1} \rm{kpc}^{-2}$ and $\sunm \rm{pc}^{-2}$, respectively. We
employ $E_{\rm SN}=10^{51}$erg, a top-heavy IMF index $\beta=1.9$, $f_c=0.63$,
$f_*=0.03\sunm^{-1}$, $\alpha_{\rm SS}=0.2$, and $\alpha_{\rm SN}=1$ (in light of
$D_{\rm SN}\sim H$) throughout the paper. We take
the outer radius $R_{\rm out}=1.0$ pc and $\mbh=10^8\sunm$, but adjust two parameters
$\epsilon$ and $\dot{M}_{\rm out}$ to demonstrate the final accretion rates of SMBHs.

\subsection{Structure of the SF disk}
Figure 2{$a_1$} shows that the inflow driven by SNexp holds at a level of
$\dotmbh=0.5\sim 2\dot{M}_{\rm crit}$ at the self-gravity radius ($R_{\rm sg}\sim 0.02$
pc) for $\epsilon=0.5$.  This clearly demonstrates that an SF disk is able to drive an
SS disk. Figure 2{$b_1$} plots $\sigsfr$ along the radial direction. We find that there is a
peak of $\sigsfr$ in the SF disks
before switching on an SS disk. This is caused by the star formation and the radial
transportation of gas that compete along radii. From the second term of Equation (7),
there is a cutoff of $\sigsfr$ and then an SF disk switches on an SS disk.
Figure 2{$c_1$} displays a smooth transition of the surface density from SF disks to SS
disks. The bumps in Figure 2$c_1$ are artificially made by the assumption for simplicity
that there is no star formation in the SS disk. This assumption roughly holds since the
temperature within $R_{\rm sg}$ could be high enough to efficiently suppress star
formation.

The right column of Figure 2 shows a dependence of the structure on $\epsilon$ for fixed
$\dotmout$. We find that $\dotmbh$ increases with $\epsilon$, which appears in the
exponential as shown in Equation (7). For $\dotmout=3\sunmyr$, the accretion rates
are about 5\%-10\% $\dotmout$ from $\epsilon=0.2-0.6$. The fraction $(1-\epsilon)$ of
the SNexp kinetic energy will be used to heat the gas of the SF disk, and then star
formation is quenched close to the middle regions of SS disks. The output of this energy
appears in the infrared bands in spectral energy distributions. Figure 3 shows the
dependence of accretion rates on the global star formation rates, defined by
$\calr=\int 2\pi \sigsfr RdR$. For a fixed $\epsilon$, we
find $\calr\propto \dotmbh^q$, where $q=1.43 - 1.56$ insensitive to $\dotmout$.

{\centering
\figurenum{4}
\includegraphics[width=2.5in,angle=-90]{fig4.ps}
\figcaption{Comparison of the present model with the observed metallicity-luminosity relation.
We use the theoretical relation $\log Z/Z_\odot=1.0+1.33\log$(\nv/\civ)\, (Hamann et al. 2002) to
convert \nv/\civ\, into $Z$ for the Shemmer et al. (2004) sample. We employ the relation
$L_{1450}\approx 2L_{5100}\approx (2/9)L_{\rm bol}$, $L_{\rm bol}\approx 9L_{5100}=\eta \dotmbh c^2$
where $\eta$ is the radiative efficiency and $L_{\rm Bol}$, $L_{1450}$,
and $L_{5100}$ are bolometric and specific luminosities at 1450\AA\, and 5100\AA, respectively.
We use $Z_0=0$, $\varepsilon_1=20$, and $\varepsilon_2=0.1,0.05,0.01$ for red, green, and blue solid
lines, respectively, whereas the dotted lines are for $\varepsilon_1=10$. The theoretical
metallicity--luminosity relation is consistent with the observed.}
}
\vglue 0.2cm

It is necessary for AGNs and quasars to switch the SF disks from the middle region of
the SS disks. We find that for lower $\dotmout$ the SF disks only support the outer
region of the SS disk.
In such a case, the SF disk may drive an advection-dominated accretion flow (ADAF) to SMBHs. 
Excellent examples of this case could be the Galactic center, some elliptical
galaxies (e.g., M87), and LINERS, where star formation exhausts most of gas forming an ADAF
as discussed by Tan \& Blackman (2005). We would like to point out that
$\beta$ used here is only an intermediate top-heavy IMF. It has been suggested that
$\beta=0.85$ (Maness et al. 2007) and
$\beta=0.45$ (Bartko et al. 2010) in the Galactic center.

\subsection{Relation Between metallicity and luminosity}
Metals are ejected through SNexp enhancing the metallicity in
the SF disk. The mass of the metal element $-i$ produced by SNexp is given by
$M_{Z_i}=\int_{M_1}^{M_2}dM_*N(M_*)\calr m_i^{\rm ej}(M_*,Z_*)$, where $m_i^{\rm ej}(M_*,Z_*)$
is the ejected mass of the element$-i$ dependent on the initial mass and metallicity of progenitor
stars (Woosley \& Weaver 1995; Gavilan et al. et al. 2005). With the $M_{Z_i}$, the total ejected
metal is then calculated by $M_Z=\sum_i M_{Z_i}$.
For simplicity, we take an approximation of the metal fraction $m_Z=M_Z/M_*\approx 0.1-0.2$
(Woosley \& Weaver 1986).

Detailed calculations of metallicity involve evolution of stellar populations
in the SF disk, advection of metal-enriched gas, and star formation, which make a strong 
radial dependence of metallicity complicated. However, we study the mean metallicity
$Z$, as the observed, in the SF disk by introducing $\varepsilon_1$ and $\varepsilon_2$.
Considering the radial dependence of metallicity, $\varepsilon_1Z$ will be swallowed by SMBHs
while $\varepsilon_2 Z$ is consumed by star formation. The lower limit of metallicity at radius
$R$ is roughly given by $Z\approx f_c\mz\dotcalms/\dotm$, and we have
$\varepsilon_1\ge Z_{\rm in}/Z\sim \dotcalms^{\rm in}\dotm/\dotcalms\dotmbh\sim \dotm/\dotmbh\sim 5$
from Figure 2($a_1$), where the (sub/super) script ``in'' means the values of parameters at the inner 
edge of the SF disk, and $\dotcalms$ is the local SF rates. On the other hand, its upper limit is
$Z_{\rm max}/Z\sim f_c\mz\calr/\dotmbh Z\sim 30/(Z/Z_\odot)$, where we use $f_c=0.6$
and $\mz=0.1$. So we have $5\le \varepsilon_1\le 30$ if the mean $Z\sim Z_\odot$. The parameter
$\varepsilon_2\sim Z_0/Z\sim 0.1$ for $Z_0=0.1Z_\odot$.

The total mass of the metal by time $t$ through integrating the net increase of metals is given by
$M_Z(t)=\int_0^t \left(Z_0\dotmout +f_c\calr \mz-\varepsilon_1Z\dotmbh-\varepsilon_2 Z\calr \right)d \tau$,
where $Z_0$ is the initial metallicity of the gas supplied from the torus. The first term in the
integration is contributed by the initial metal of supplied gas, the second is contributed by SNexp,
the third is advected into SMBH and the last one is the consumption of metals during star formation.
The total gas of the SF disk is given by
$M_{\rm gas}(t)=\int_0^{R_{\rm out}}2\pi \Sigma_{\rm gas}(t,R)RdR$.
We obtain the metallicity at time $t$ by its definition of $Z(t)=M_Z(t)/M_{\rm gas}(t)$
\begin{equation}
Z(t)M_{\rm gas}(t)={\displaystyle{\int_0^t} \left(Z_0\dotmout +f_c\calr \mz-
                   \varepsilon_1Z\dotmbh-\varepsilon_2 Z\calr \right)d \tau}.
\end{equation}
This is an integral-differential equation, which is easy to understand in light of $Z$
as an integral parameter with time. For a steady SF disk, its total mass is a constant,
but metals are increasing. After some algebraic manipulations, we have
its solution
\begin{equation}
Z(t)=Z_{\rm max}+\left(Z_0-Z_{\rm max}\right)\exp\left(-\frac{t}{t_Z}\right),
\end{equation}
where $Z_{\rm max}=(Z_0\dot{M}_{\rm out}+f_c\calr m_z)
/(\varepsilon_1\dotmbh+\varepsilon_2\calr)$, $t_Z=M_{\rm gas}/(\varepsilon_1\dotmbh+\varepsilon_2\calr)$,
and $Z(0)=Z_0$ is assumed initially.  For a time long enough,
the last term will tend to zero, we have
\begin{equation}
Z_{\rm max}=Z_0(1-\Lambda)+f_c(\mz-Z_0)\Lambda
\approx \frac{c_0f_c\mz}{\varepsilon_1+\varepsilon_2c_0\dotmbh^{0.4}}\dotmbh^{0.4},
\end{equation}
where $\Lambda=\calr/(\varepsilon_1\dotmbh+\varepsilon_2 \calr)$,
$\dot{M}_{\rm out}=(1-f_c)\calr+\dotmbh$ is used,
$\dotmbh$ is in units of $\sunmyr$, $\calr=c_0\dotmbh^{1.4}$,
and $c_0=43.03$ insensitive to $\mbh$ for $\epsilon=0.5$. The approximation is valid
only for $Z_0\ll \mz$. We assume for a comparison with observational data that most
of quasars reach the state with maximum metallicity, but the episodic age of quasars
can be estimated from Equation (14) if the metallicity is measured accurately enough.

Figure 4 shows a comparison of the model with data. Not only
the metal-rich phenomenon in quasars is explained, but also the {\textit{Z--L}}
relation is reproduced by the present model. The large scatters of the
{\textit{Z--L}} relation imply different $\varepsilon_1$, $\varepsilon_2$, and $c_0$ individually.

\subsection{Compact nuclear star clusters}
The present model may predict the formation of nuclear star clusters commonly found in
local galaxies. They are 1-2 orders of magnitude brighter than globular clusters
(C\^ot\'e et al. 2006), have extended star formation histories (Rossa et al. 2006),
complex morphologies (Seth et al. 2006), and follow the relation similar to the Magorrian
relation (Ferrarese et al. 2006). Though the SF disk discussed here is only on a scale of
1 pc, SF is ongoing inside the torus at a few 10 pc scale (Collin \& Zahn 1999), which
is invisible due to dust extinction. Since
AGN types are irrelevant to orientations of their host
galaxies (e.g., Munoz Marin et al. 2007), SMBHs are randomly fed by the dusty torus.
This is evidence for random accretion onto the SMBHs, which is derived
from the $\eta$-equation (Wang et al. 2009a) and is confirmed by
calculations based on the semi-analytical theory of mergers (Li et al. 2010). One stellar
belt with random direction composed of old stars is left after one episodic activity accordingly.
The appearance of the nuclear star cluster could
be a natural consequence of the multiple episodes of SMBH activities (Wang et al. 2008;
Wang et al. 2009a). The clusters are
thus tightly related to the central SMBHs in the present model. Figure 3 suggests that a
mass ratio of SMBH and a nuclear star cluster is roughly 0.01-0.1 depending on $\epsilon$ and
$\beta$ (i.e., star formation history), consistent with the observations (Seth et al. 2008).
Detailed comparison with observed properties of the clusters, such as radial-dependent
metallicity, SF history, will
provide further clue to understand physics in galactic centers.

All the emission from the SF disk will be thermalized
with a typical temperature of
$T_{\rm SF}\sim \left(L_{\rm SN}/2\pi R_{\rm SF}^2\sigma_{\rm SB}\right)^{1/4}
\sim 2000~L_{\rm SN,43}^{1/4}$K in the near-infrared band, where $\sigma_{\rm SB}$
is the Stefan--Boltzman constant,
$L_{\rm SN,43}=L_{\rm SN}/10^{43}{\rm erg~s^{-1}}$ is the SNexp luminosity. It
then contributes a fraction to the observed infrared emissions in quasars, which is
worth investigating further.

\section{Conclusions and discussions}
We show that an SF disk supplied by the dusty torus is able to support an SS disk 
around SMBHs in AGNs and quasars. Such a model
naturally produces the observed relationship between metallicity and luminosity.
As a natural consequence of the multiple episodes of activity
through random accretion onto the holes, a nuclear star cluster will be formed from
the remnants of stellar disks. The present model would provide a unified explanation
of feeding SMBHs, production
of metallicity, and formation of a nuclear star cluster.

We stress that the present paper deals with the stationary structure of SF disks 
switching to SS disks. A time-dependent model will
give the evolution of AGNs driven by the SF disks for the
complicated AGN--starburst connection.

\acknowledgements The anonymous referee is thanked for a helpful report. H. Netzer is
acknowledged for interesting discussions. We appreciate the stimulating discussions
among the members of the IHEP AGN group. The research is supported by NSFC-10733010
and 10821061, CAS-KJCX2-YW-T03, and 973 project (2009CB824800).

\end{document}